\documentclass[11pt,usenames,dvipsnames]{article}

\usepackage[normalem]{ulem}
\usepackage[swedish,english]{babel}

\usepackage{color}
\usepackage{xcolor}
\usepackage{amssymb}
\usepackage{amsmath}
\usepackage{ae}
\usepackage{slashed}
\usepackage{graphics}
\usepackage{graphicx}
\usepackage{url}
\usepackage{hyperref}
\hypersetup{linktocpage}

\usepackage{cite}
\usepackage{times}
\usepackage{xfrac}

\usepackage{fancyhdr}
\usepackage{enumerate}
\def\be{\begin{equation}}
\def\ee{\end{equation}}

\fancypagestyle{jourref}{\fancyhf{}\fancyfoot[L]{\it 
This version of the article has been accepted for publication, after peer review, but is not the Version of Record and does not reflect post-acceptance improvements, or any corrections. 
The Version of Record of this article is published in Foundations of Physics, and is available online at \href{https://doi.org/10.1007/s10701-022-00667-6}{https://doi.org/10.1007/s10701-022-00667-6}.}}

\usepackage{tikz}
\numberwithin{equation}{section}
\usepackage{tcolorbox}
\usepackage{bbold}
\usepackage{enumitem}

\begin{document}
\selectlanguage{english}
\frenchspacing
\pagenumbering{roman}
\begin{center}
\null\vspace{\stretch{1}}

{ \Large {\bf
Building spacetime from effective interactions\\ between quantum fluctuations
}}
\\
\vspace{1cm}
Anna Karlsson
\vspace{1cm}

{\it
Institute for Theoretical Physics, Utrecht University,\\
Princetonplein 5, 3584 CC Utrecht, The Netherlands}
\vspace{0.5cm}

{\it
Department of Mathematical Sciences, Chalmers University of Technology\\ 
and the University of Gothenburg, SE-412 96 Gothenburg, Sweden}
\vspace{1.6cm}
\end{center}

\begin{abstract}
We describe how a model of effective interactions between quantum fluctuations under certain assumptions can be constructed in a way so that the large-scale limit gives an effective theory that matches general relativity in vacuum regions. This is an investigation of a possible scenario of spacetime emergence from quantum interactions directly in the spacetime, and of how effective quantum behaviour might provide a useful link between detailed properties of quantum interactions and general relativity. The quantum fluctuations are assumed to entangle sufficiently for a cohesive spacetime to form, so that their effective properties can be described relative to a $D$-dimensional reference frame. To obtain the desired features of a smooth metric with a vanishing Ricci tensor, the quantum fluctuations are modelled as Gaussian probability distributions, with a shape set relative to the interactions coming from the surroundings. At small scales, the propagation through the spacetime is modelled by a Gaussian random walk.
\end{abstract}

\vspace{\stretch{3}}
\thispagestyle{jourref}
\newpage

\pagenumbering{arabic}

\section{Introduction}\label{s.intro}
Quantum gravity remains a puzzle despite numerous attempts at gaining insight into what it is characterised by. The best current understanding of it comes from analyses using the gauge/gravity duality \cite{Maldacena:1997re,Witten:1998qj,Gubser:1998bc}, which indicate that entanglement should play a key role in the physics. Moreover, the gauge/gravity duality belongs to a specific type of scenario for quantum gravity, where the spacetime in general relativity (GR) is emergent from quantum physics. In general, the two key questions of spacetime emergence are what the relevant processes at the quantum level are, and how the related dynamics gives rise to an emergent spacetime that is characterised by Einstein's field equations.

In a search for a quantum model that at large scales captures general GR spacetimes --- not just those with suitable gauge duals --- a key role of entanglement for spacetime emergence leads towards that the relevant physics concerns interactions between quantum particles, in a many-body system setup. A simple ansatz for general spacetime emergence then is the scenario where interactions between quantum fluctuations, present directly in the vacuum regions of spacetime, give rise to GR (in vacuum regions) at large scales. This type of emergence is analogous to that of effective theories in many-body physics, e.g. how the effective theory of temperature emerges from the more detailed dynamics of the kinetic energy of molecules in an ideal gas.

We present an effective quantum model for interactions between quantum fluctuations, which gives GR in vacuum regions as an effective theory in the large-scale limit. It represents a possible scenario for the simple ansatz for spacetime emergence described right above, and might provide a link between the individual interactions present between quantum particles and the effective theory of GR. At the quantum level, the model is effective in that it assumes that the correlations between the quantum fluctuations are sufficient to give rise to an effective spacetime --- so that the behaviour of the particles can be described relative to a $D$-dimensional reference frame --- and in that it only details an effective, conjectured behaviour of the quantum fluctuations. In particular, it builds on the results of \cite{Karlsson:2019avt}, where the same ansatz was analysed for flat spacetimes (characterised by a vanishing Riemann tensor), and where the interaction rate profiles of the quantum fluctuations were found to be required to be Gaussian functions. The effective behaviour of the quantum fluctuations which is used in the effective quantum model, is a possible scenario for their statistical interactions with their surroundings.

The key features of the effective quantum model we analyse are the following. In the model, the spacetime position of each quantum fluctuation is given by a Gaussian probability distribution, that is set in relation to the incoming interaction it receives from its surroundings. In combination with the expected momentum of the particle, $P_o^\mu$, the Gaussian distribution specifies the propagation of the particle in terms of a random walk. This is a simplified picture of (a possible scenario for) the dynamics of the quantum fluctuations, which turns out to be useful. Based on the model, a metric can be defined, and the Ricci tensor can be shown to vanish. In addition, the interactions can be analysed in more detail from the perspective of what is required of them for said effective behaviour to arise. In this sense, the model we analyse hopefully might provide a step towards understanding how spacetime can emerge from quantum interactions. It might provide a useful link between the detailed interactions and the effective large-scale theory.

\subsection{Motivation for the ansatz, and the Gaussian distributions}\label{s.mot}
We focus on a specific scenario of spacetime emergence, where the effective theory (GR in vacuum regions) emerges in the large-scale limit of an effective quantum model. The scenario is built on the
\begin{itemize}[leftmargin=22mm]
\item[Ansatz:] GR spacetime has an origin in interactions between quantum fluctuations.
\end{itemize}
which relates to the connection between spacetime and entanglement observed in gauge/gravity duality analyses in the sense that a key role of entanglement implies that quantum interactions are central to understanding the quantum physics. While the quantum interactions in the gauge/gravity duality take place on the boundary, a model for general spacetimes would be easier to obtain from processes directly in the spacetime. A candidate process for that is interactions between quantum fluctuations, which are naturally present in the vacuum. The possibilities such a scenario presents constitutes a first motivation for considering the ansatz above.

In addition, the compatibility of the ansatz with spacetime physics was analysed for flat spacetimes in \cite{Karlsson:2019avt}. Spacetimes with a vanishing Riemann tensor are typically considered trivial, but an interesting aspect of the ansatz above is that an origin in interactions present in the spacetime infers an origin in how information is exchanged, and a sensitivity to how information flows through the spacetime. In that sense, the configuration of a flat spacetime would be non-trivial. In the presence of objects that restrict information exchange, e.g. a typical slit set-up for particle diffraction, the present ansatz for spacetime emergence comes with specific flat spacetime configurations that, while characterised by a vanishing Riemann tensor, are set relative to how said objects extend in the spacetime. In the slit example, the set-up blocking the light can be approximated as insulating with respect to information exchange\footnote{I.e. representing a system that does not transmit the interactions from the quantum fluctuations.}, but massless in terms of its negligible impact on the spacetime at the scales under consideration. Effectively, those objects represent boundary conditions to the vacuum regions.

The interesting feature of the non-trivial flat spacetime configurations that come out of the slit set-ups under the current ansatz, is that they have the same symmetries as the diffraction patterns associated with the boundary conditions. Effectively, said spacetime configurations could give rise to effects attributed to the wave-particle duality. For details on how this works, we refer to \cite{Karlsson:2019avt}. In the present text, we do not discuss the relevance of the configuration of the spacetime for how information flows, and how specific metrics (relative to the boundary conditions) capture that. However, the feature that the ansatz could provide a mechanism for the wave-particle duality is interesting. In models for new physics, it is desirable to find predictions that can be experimentally verified, as proof of that the models are correct. While there is no evidence for a scenario where said wave-particle duality effects arise from the flat spacetime configurations discussed above, the coinciding features of the ansatz and the observed physics (i.e. the wave-particle duality) nevertheless lends support to the ansatz, and to the scenario of spacetime emergence that we analyse.

The results of \cite{Karlsson:2019avt} also include observations on what the interactions between the quantum fluctuations would have to be characterised by to give rise to a flat spacetime. In specific, the rate at which a quantum fluctuation interacts with its surrounding would have to have a Gaussian fall-off, in spacetime. The identified reasons for this are as follows. {\it (i)} The flat spacetime configurations are sensitive to boundary conditions modulo Gaussian profiles, and without an imposed scale (as in edge diffraction) the fall-off of the non-trivial configuration is Gaussian. {\it (ii)} The interaction rate profiles would provide basis functions to the spacetime, and since the metric is smooth, the basis functions ought to be Gaussian. {\it (iii)} Gaussian interaction profiles fit with that consecutive interactions (by necessity products) add up in terms of length, as given by the line element and as illustrated by Pythagoras' theorem for $e^{a^2}e^{b^2}=e^{c^2}$. For more details on this, we refer to \cite{Karlsson:2019avt}. 

In the effective quantum model, we use the result that the interaction profiles of the quantum fluctuations are given by Gaussians, but with a modified interpretation. Effectively, the position of a quantum particle is given by where it initiates an interaction. Hence the Gaussian function equally can be interpreted as detailing a probability distribution for the spacetime position of the quantum fluctuation. This is how we transition from Gaussian interaction rates to Gaussian probability distributions for the positions of the quantum fluctuations. The latter are far more useful in an effective quantum model.

\subsection{The basics of how the effective model works}\label{s.base}
When modelling small-scale physics of GR, there are two modes of approach: {\it (i)} to impose GR restrictions at small scales, and {\it (ii)} to only require that the GR properties arise at large scales. The present ansatz is of the second kind. Consequently, the small-scale physics (the dynamics of the interactions) in the model is not in GR, and not constrained by locality etc. The particles can just as well be thought of as existing at a single point, but it is useful to encode their probability of interacting with each other through positioning them relative to a reference frame; this is a convenient visualisation of the particle dynamics. That same positioning also makes it straightforward to recover GR in the large-scale limit. It is {\it only} in the large-scale limit that GR properties arise.

The model ansatz is that the probability distribution for the interaction initiated by a particle, in a local reference frame, is given by a Gaussian, for the reasons given in \S\ref{s.mot}. At small scales, this means that the particle performs a random walk relative to the reference frame. These random features are not detectible in the large-scale limit, i.e. in the theory agreeing with GR. It is important to keep in mind that while the small-scale theory is an ansatz for spacetime dynamics at small-scales, it is not GR. The model specifies both the position and the momentum of each particle, as detailed in \S\ref{s.Gprob}, and the interactions are allowed to display spurious event of e.g. $v>c$, as visible in eq. \eqref{eq.Ppos} and \eqref{eq.Pmom}. The important thing is that such events are sufficiently suppressed to allow for an average behaviour consistent with GR to form at large scales, much how the chaotic behaviour of molecules in a fluid on average conforms to fluid dynamics, at large scales. The suppression in the present model is strong enough: the deviations are exponentially suppressed, and the standard deviations of the distributions are of order the Planck scale, as described in \S\ref{s.Gprob}. Note also that in models where both the position and momentum distributions are specified, it is required to accommodate large deviations in the same way as by $\sigma_{x}\sigma_{p_x}\geq\hbar/2$ by Heisenberg's uncertainty principle. In our model, we simply fit the momentum distribution of a particle to its given position distribution, so that the momentum distribution is the minimal distribution allowed by Heisenberg's uncertainty principle.

At large scales only the average of the small-scale dynamics survive. We do not treat the individual interactions, but only their probability distributions and how those distributions evolve over the (short) particle lifetime. In specific, we show that the acceleration of the expected position of a particle is given by the geodesic equation. Since the distributions are Gaussian, their time evolution is well-known, and in \S\ref{s.rWalk} we use those properties to show how a gradient in interaction rate (relative to the reference frame) affects the average movement of a particle. Here, note that the model has a key difference from standard Brownian motion: the standard deviation of each Gaussian distribution is set relative to the incoming interaction from the particle's environment (see \S\ref{s.eqmod} for more details) instead of constituting a constant in a Minkowski reference frame. This ensures diffeomorphism invariance, and causes the particle propagation to deviate from standard Brownian motion. In the model, the average step length is a function of the particle's position in spacetime. The effect is most easily pictured in terms of pressure. If the interaction rate (from the environment) were pressure, the average step length would be set relative to a fix pressure, and a gradient in pressure would cause a bias towards lower pressure, giving a shift in the expected position of the particle that corresponds to acceleration. In our ansatz, $g_{tt}$ governs the interaction rate, and the model encodes a bias towards lower $g_{tt}$. In \S\ref{s.geo} we detail how the acceleration of the expected position of a particle is described by the geodesic equation. By construction, our modification of the random walk causes an acceleration as described in \eqref{eq.peq}. If one desires to think of the random walk in terms of standard Brownian motion, the effect is the same as if the standard propagation were to take place on the surface the spacetime describes when embedded in Minkowski space; the fact that the surface is curved creates additional acceleration relative to the reference frame as the position distribution of the particle spreads over the surface. We explain how the bias to the random walk works in \S\ref{s.rWalk}.

Finally, the model ansatz is that the particles, through their position distributions, define basis functions for the spacetime. The average variance of the $D$-dimensional Gaussians defines the metric $g_{\mu\nu}$ as described in \S\ref{s.met}, and in \S\ref{s.Ricci} we identify the conditions for $R_{\mu\nu}=0$. Also note that since we discuss {\it quantum} interactions, the entire discussion effectively is about entanglement.

\subsection{Summary of the model: the main assumptions and results}
We describe how a many-body system of quantum fluctuations can be characterised by effective dynamics at the quantum level, which at large scales provides dynamics coinciding with GR. The main assumptions that go into the construction of the effective quantum model are:
\begin{itemize}[leftmargin=0.5cm]
\item {\it The key model features of the quantum fluctuations are conserved under creation/annihilation of the quantum fluctuations.}

A necessary requirement for any model building on quantum fluctuations.

\item {\it The properties of the quantum fluctuations can be described relative to a $D$-dimensional reference frame. }

This assumption provides a straightforward way to connect effective properties at the quantum level to large-scale physics. It is equivalent to assuming that the interactions between the quantum fluctuations are sufficient for the particles to become entangled, and agree on distance and orientation enough for a cohesive spacetime to form. Note that the inter-particle dynamics is internal only: the quantum physics described is independent of the choice of reference frame. Eventually, it is relevant to lift this assumption to understand the quantum interactions properly, as well as what happens when the effective (GR) theory breaks down.

\item {\it A Gaussian profile for the interaction rates of the quantum fluctuations.}

The motivation for this is described at the end of \S\ref{s.mot}. It infers a Gaussian probability distribution for the position of a quantum fluctuation, which is compatible with Heisenberg's uncertainty principle, and with a smooth metric.

\item {\it The Gaussian distribution also describes a random walk of each quantum fluctuation.}

This provides a mechanism for how the propagation of each quantum fluctuation obeys the geodesic equation, and for $R_{\mu\nu}=0$. Each step of the particle propagation is set by two displacements, one from the Gaussian distribution and one by the average momentum $P_o^\mu$, without a preferred order of the displacements.

\end{itemize}
The main result is that, based on the ansatz that GR has an origin in interactions between quantum fluctuations, an effective quantum model can be constructed that gives a GR metric with $R_{\mu\nu}=0$ in the large-scale limit. The effective model treats each quantum fluctuation as a Gaussian probability distribution in spacetime, with its shape set relative to the interactions the particle receives from its surroundings. Relative to an embedding in Minkowski space, these extended objects constitute basis functions (of equal weight) to a surface that describes the spacetime. The quantum physical features are only dependent on the internal dynamics, and each particle performs a random walk through the small-scale rendition of the spacetime. The random walk in turn is constructed so that it defines $R_{\mu\nu}=0$, and at large scales the particle propagation by default follows the geodesic equation, which shows as an artefact of how the spacetime is represented in the reference frame.

The focus of the present model is vacuum regions of spacetime. Anything out of the vacuum is modelled through boundary conditions. Objects passing through the spacetime are assumed to interact with the quantum fluctuations, and move relative to the internal dynamics they define. The interactions between the quantum fluctuations are treated effectively, and modelled by the Gaussian distributions and the random walk of the particles described above. A brief comment on electric fields, and on the precise interactions and properties of the quantum fluctuations as well as on how the spacetime dimensions actually emerge, can be found in the appendix. Of course, it is the precise properties of the quantum interactions that need to be understood for a model of quantum gravity in a scenario of emergent spacetime. The present model is a rough, possible set-up that might be a useful link between the detailed quantum physics and the classical theory. Overall, the model has several parallels to many-body physics.

\subsection{Comparison with other scenarios for spacetime emergence \& outline}
The model we analyse concerns a scenario for spacetime emergence that is special in the sense that the emergence comes from effective quantum interactions, directly in the spacetime theory. The emergence of the large-scale theory most closely resembles that of effective theories in many-body physics. However, the emergence from quantum interactions also means that it should be possible to connect, or show similarities of, the model scenario to the spacetime emergence that is present in the gauge/gravity duality, and in approaches to spacetime emergence that use the gauge/gravity duality, such as the observations made in \cite{VanRaamsdonk:2010pw}, constructions using tensor networks \cite{Swingle:2009bg,Swingle:2012wq} and analyses of matrix models. While the present model builds on effective quantum processes instead of detailed interactions, as is done in the gauge/gravity duality approaches, the effective behaviour imposed on the quantum fluctuations does require specific interaction properties. It is with a more detailed picture of those interactions that it might be possible to identify similarities with the gauge/gravity duality approaches. Superficial similarities are readily present, such as that the small-scale physics of the effective model is independent of the choice of reference frame (as long as the reference frame can capture the full internal dynamics), which means that a dimensional reduction of the reference frame should be possible, at least for special cases. If e.g. a $2D$ effective quantum model were to be reduced in that fashion, the Gaussian distributions in the removed spatial direction would translate into chains of particles correlated by Gaussian probability distributions. This has similarities with the Sachdev--Ye--Kitaev (SYK) model \cite{Sachdev:1992fk,Kitaevx}, which in a certain limit is dual to $2D$ gravity. 

It is good to note that relativistic random walks is a subject on its own. For an overview, see e.g. \cite{dunkel}. As specified in \S\ref{s.base}, we do not treat random walks in GR, but to do so comes with several complications, including how to find a replacement for the Gaussian distribution that does not violate locality etc., so that the random walk conforms with relativistic Brownian motion. For relativistic random walks, it is also best if the averages can be calculated in a manifestly covariant way. We avoid these difficulties through not imposing GR restrictions on the small-scale dynamics.

We begin with a presentation of the effective quantum model in \S\ref{s.eqmod}. A thorough analysis of the Gaussian probability distributions is given in \S\ref{s.Gprob}. Following that, we describe how the metric arises in \S\ref{s.met}, and how the random walk of the quantum fluctuations gives a propagation described by the geodesic equation, in \S\ref{s.rWalk}-\ref{s.geo}. Finally, we show how the model sets $R_{\mu\nu}=0$ in \S\ref{s.qRicci}, and how boundary conditions to the vacuum regions give rise to different types of spacetime solutions, in \S\ref{s.bound}, before we end with a summary and an outlook.

\section{The quantum model}\label{s.qmod}
Our ansatz for spacetime emergence is that quantum interactions directly in the spacetime give rise to GR in a large-scale limit. Earlier analyses of flat spacetime \cite{Karlsson:2019avt} have shown that a candidate model, based on this ansatz, is characterised by that each quantum fluctuation has an interaction rate with its surroundings that follows a Gaussian fall-off. These Gaussian profiles then figure as basis functions of the spacetime, which we get back to in \S\ref{s.rise}.

In this section, we build on the earlier observations and construct an effective quantum model for vacuum regions, which at large scales gives GR. This effective model uses the assumption that the interactions between the quantum fluctuations are frequent enough (and under otherwise benign conditions) to give rise to a cohesive spacetime, so that it makes sense to use a reference frame\footnote{We use the shorthand notation $\{x^\mu\}$ for a set of spacetime coordinates.} $\{x^\mu\}$ to describe where the quantum fluctuations are located relative to one another. In this setting, the quantum fluctuations figure as Gaussian probability distributions in spacetime, which perform random walks through the spacetime.

The effective quantum model is the centre piece of the present text. An effective model relative to $\{x^\mu\}$ provides the most straightforward way to show that the effective theory at large scales corresponds to GR. Of course, at the level of the quantum interactions, the model includes more details, but over larger scales --- at a scale where quantum features are detectible, but single quantum interactions are not relevant --- it is sufficient to treat the quantum fluctuations as said Gaussian distributions in spacetime, relative to some $\{x^\mu\}$. Note that we only treat the interactions at said effective level, in this text. A short discussion on the associated physics at the level of individual quantum interactions can be found in the appendix.

\subsection{The effective quantum model}\label{s.eqmod}
Consider the following set-up. A quantum fluctuation, i.e. a particle that temporarily is present in the vacuum, interacts with other quantum fluctuations that in some sense belong to its environment. The particle itself is characterised by that it can interact with the other particles, in a series of events that belong to a certain number of time-like dimensions (where events are logged in sequence) and by that it can receive incoming interactions relative to a unit sphere $S^{d-1}$. This latter condition is equivalent to that the particle carries physical properties that depend on the $S^{d-1}$, such as spin 1/2 particles do for $d=3$ through the orientation of their spin. Upon receiving incoming interaction, the particle effectively creates the perception of a unit $S^{d-1}$ relative to its surroundings: based on the spatial angle of the incoming interaction (relative to its $d$-dimensional physical properties) and the frequency of the incoming interaction (as a function of spatial angle) it effectively creates a $S^{d-1}$ perception of the surroundings, where the interaction rate is constant over the sphere. This sets a local definition of spatial length and orientation. Moreover, the rate of incoming interaction (number of events logged) also serve as a measure of the time that has passed\footnote{In principle, there can be several logs of this kind, corresponding to multiple time-like directions.}. Based on these two constructions of space and time, the particle then initiates an interaction with its surroundings, with a probability distribution given by a Gaussian profile around a unit $S^{D-1}$, which in the rest frame of the particle coincides with the $S^{d-1}$ constructed by the particle, and in addition encodes a Gaussian probability distribution around exactly where the interaction takes place in the time-like event log. After the interaction has been initiated, the process starts over again.

The above describes a process for how a quantum particle can construct a spacetime-like perception of its `surroundings' based on the incoming interactions it receives from other particles. Now, there are certainly conditions that need to be met for a set of particles to give rise to a cohesive structure where this `spacetime-like picture' extends between several particles to the point that a large-scale entity characterised by spacetime properties is formed. We discuss a few of those conditions in the appendix. However, it is clear that under sufficiently beneficial conditions, the spacetime quality of the individual particles can be linked to form a spacetime at large scales. Under the assumption that this is the case, the interaction process described above can be described relative to a reference frame $\{x^\mu\}$.

Relative to the $\{x^\mu\}$, the quantum fluctuation interacts at a point in spacetime, $x_o^\rho$. It then receives incoming interaction from particles nearby. The $S^{D-1}$ it constructs corresponds to a local definition of time and spatial length in the spacetime, which is equivalent to the presence of a reference frame where the line element $ds$ is given by
\be\label{eq.locr}
ds^2\bigg|_{x^\rho=x_{o+}^\rho}=\eta_{\mu\nu}dx^\mu dx^\nu\,.
\ee
Here, $x_{o+}^\rho$ denotes the next, expected interaction point after $x_{o}^\rho$ (the input the quantum fluctuation receives after $x_o^\rho$ decides its perception of spacetime at $x_{o+}^\mu$). We use $c=1$, and time is given relative to the frequency of the particle; the time between two initiated interactions is one unit of time. The line element itself reflects both a difference between the spacetime directions (time-like vs space-like) and how information (i.e. interaction) is sent out from, and received by, the particle: with a vanishing line element that requires propagation in time and space to go hand in hand. 

Finally, relative to the $\{x^\mu\}$, the next interaction point (in spacetime) of the particle is given by a displacement set by the momentum $P_o^\mu$ of the particle ($x_o^\rho\rightarrow x_{o+}^\rho$) in combination with a displacement set by a Gaussian probability distribution around $x_{o+}^\rho$, in the reference frame of \eqref{eq.locr}. There is no preferred order to the two displacements, and the Gaussian distribution fills two functions. It describes the spacetime propagation of the particle in terms of a random walk (on top of the movement required by the momentum), and in doing so it also gives a probability distribution for where the particle is located in spacetime (at the next step). 

Effectively, for a model that does not concern itself with the individual interactions of the quantum particles, each quantum fluctuations can be treated as a Gaussian probability distribution in spacetime, which propagates through the spacetime in a random walk set by the interactions it shares with the quantum fluctuations in its environment. This provides an effective description of some of the quantum physics of the spacetime, given the current ansatz for spacetime emergence.

In the above, we introduced a momentum $P_o^\mu$ for the quantum fluctuation, in addition to its spacetime position. In the rest frame of the particle, characterised by \eqref{eq.locr}, the only non-zero components of $P_o^\mu$ are the temporal ones, which give rise to a displacement of one unit (for each time-like direction the particle propagates in) between each interaction. With a change of reference frame, $P_o^\mu$ can be altered to include spatial displacements as well. However, note that at the quantum level the only {\it physical} processes are given relative to the $S^{D-1}$s of the quantum particles --- not relative to the reference frame.

In the effective quantum model, both the position in spacetime and the momentum of a quantum fluctuation are given by probability distributions; the uncertainty in spacetime position by necessity infers an uncertainty in momentum, which will be further discussed in \S\ref{s.Gprob}. The particle interactions will also come with a transfer of momentum between particles, while the total momentum (given by the $P_o^\mu$s) remains conserved. With respect to this, the interactions will figure as typical collisions between particles, resulting in the same type of equilibration of momentum among particles in a spacetime volume element as is characteristic of other many-body systems.
\\\\
The set-up presented above specifies the effective quantum model of this text. The key feature of the model is the Gaussian probability distribution for the position of a quantum particle, which denotes the probability for where in spacetime the particle initiates an interaction with its surroundings. After each interaction, the particle receives incoming interaction and builds a new probability distribution, for its next step, based on that information. The next step is then set by the $P_o^\mu$ of the particle, in combination with a random draw from the probability distribution.

Note that in addition to the description above, the effective model includes an assumption of a suitable underlying creation and annihilation process of the quantum fluctuations in the vacuum, so that the properties of the quantum fluctuations are retained over time. Having described the set-up of the effective model, we will now discuss the Gaussian probability distributions further in \S\ref{s.Gprob}, and the properties of the random walk in \S\ref{s.rWalk}.

\subsection{The Gaussian probability distributions of the quantum fluctuations}\label{s.Gprob}
The effective quantum model includes a Gaussian probability distribution for the spacetime position of a quantum fluctuation, relative to the reference frame characterised by \eqref{eq.locr}. For simplicity, we here denote that reference frame by $\{\tilde x^\mu\}$. A specification of the Gaussian distribution in question requires the scale of the fall-off to be specified. In\footnote{It is good to note a few things about \eqref{eq.p2}. First, we refer to the $\sigma$ that the quantum fluctuation is characterised by at interaction with other particles in the spacetime. In an experiment, the $\sigma$ for the spacetime position of a particle can be narrowed down at the price of an increased uncertainty in the momentum of said particle (and vice versa) but that constitutes a different kind of process. Second, the $\tilde x_o^\mu$ in \eqref{eq.p2} is the $x_{o+}^\mu$ of \eqref{eq.locr}.}
\be\label{eq.p2}
(\sigma \sqrt{2\pi})^{-D}e^{-\frac{|\eta_{\mu\nu}|\tilde\xi^\mu\tilde\xi^\nu}{2\sigma^2}}\,,\quad \tilde\xi^\mu=\tilde x^\mu-\tilde x_o^\mu\,,
\ee
a value for the standard deviation ($\sigma$) of the probability distribution relative to the unit length in $\{\tilde x^\mu\}$ (set by the $\eta_{\mu\nu}$) needs to be identified. In the below, we make a choice of $\sigma^{-1}=\sqrt{2\pi}$ in the length scale ($l_q$) of $\{\tilde x^\mu\}$. This gives a nice overall coefficient, but also specifies the possibility for the particle to not take a step forward in time at the next point of interaction as a $2.5\,\sigma$ event. A unit step back in time would be a $5\,\sigma$ event. Another suitable choice would be $\sigma=l_p/\sqrt{2}$ (where $l_p$ is the Planck length) which constitutes the minimal uncertainty required by Heisenberg's uncertainty principle, in a setting where both the uncertainty in spacetime position and momentum are simultaneously minimised (with $\sigma_x\sigma_{p_x}=\hbar/2$). We will get back to requirements by Heisenberg's uncertainty principle below. A $\sigma$ in \eqref{eq.p2} of order the Planck length would also fit with that the quantum nature of the spacetime should become relevant at the Planck scale, which in the present quantum model means that individual Gaussian distributions should start to be discernible at that scale. This exposure of the Gaussians, really of the basis functions of the spacetime, must take place in time as well as in space, which means that the unit length in time\footnote{Note that with $c=1$, we have $l_p=t_p$ etc.}, i.e. the average time between interactions, must be of order the Planck length also. It is also relevant to note that a quantum fluctuation cannot be present in the spacetime for very long, during which time it must interact with its environment several times for a cohesive spacetime to be sustained. Consequently, it is reasonable to have
\be
\sigma\sim l_q\sim l_p\,.
\ee
For example, a possible scenario is $\sigma=l_q/\sqrt{2\pi}=l_p/\sqrt{2}$. That said, the $\sigma$ of the Gaussian probability distribution is not decisively fixed. In the below, we proceed with making our illustrations using $\sigma^{-1}=\sqrt{2\pi}$ in $\{\tilde x^\mu\}$. It would be straightforward to introduce a different $\sigma$ in those calculations. Importantly, an alteration of the $\sigma$ does not change any of the observations made outside this section, \S\ref{s.Gprob}. It is a scale in the overlap (correlation) between different quantum fluctuations and thus relates to how frequently they interact (equivalently, how close they are), which in turn connects back to properties of the individual interactions and requirements for a cohesive spacetime to form.

With $\sigma^{-1}=\sqrt{2\pi}$ in $\{\tilde x^\mu\}$, we get that the probability distribution for the spacetime position of an individual quantum fluctuation is given by
\be\label{eq.p1}
e^{-\pi|\eta_{\mu\nu}|\tilde\xi^\mu\tilde\xi^\nu}\,,\quad \tilde\xi^\mu=\tilde x^\mu-\tilde x_o^\mu\,,
\ee
with the total probability
\be
1=\int d\tilde\xi^D e^{-\pi|\eta_{\mu\nu}|\tilde\xi^\mu\tilde \xi^\nu}\,.
\ee
This setup in relation to the $S^{D-1}$ frame of the particle is however just a special case of how the probability distribution appears in a general reference frame $\{x^\mu\}$. For the physics to be captured correctly, the particle interactions described by \eqref{eq.locr} must be left invariant under the coordinate transformation, i.e.
\be\label{eq.refch}
\eta_{\mu\nu}d\tilde x^\nu d\tilde x^\nu=\mathsf{g}_{\mu\nu}dx^\mu dx^\nu\,.
\ee
Here, we have introduced $\mathsf{g}_{\mu\nu}$ to denote the local definition of spacetime that is given by a quantum fluctuation, relative to $\{x^\mu\}$. In general, a coordinate transformation obeying \eqref{eq.refch} deforms the probability distribution. If the resulting $\mathsf{g}_{\mu\nu}$ is a constant, the probability distribution remains a Gaussian distribution with
\be\label{eq.Ppos}
P(x^\rho)=\sqrt{|\mathsf{g}|}e^{-\pi\mathsf{g}_{\mu\nu}\xi^\mu\xi^\nu}\bigg|_{\xi^u\rightarrow i \xi^u}\,,\quad \xi^\mu=x^\mu-x_o^\mu\,,\quad \mu=(u,i)\,: \eta_{uu}<0\,, \quad \mathsf{g}_{u\mu}\stackrel{u\neq\mu}{=}0\,,
\ee
in a co-moving reference frame, and with $\mathsf{g}=\det(\mathsf{g}_{\mu\nu})$. The total probability is given by
\be
1=\int d\xi^D \sqrt{|\mathsf{g}|}\left(e^{-\pi \mathsf{g}_{\mu\nu} \xi^\mu\xi^\nu}\bigg|_{\xi^u\rightarrow i \xi^u}\right)\,.
\ee

In addition to the effective Gaussian probability distribution, this interpretation of the local interactions of a particle also provides a way to single out the effective frequency at which the particle interacts with its environment, i.e. on average relative to a chosen reference frame $\{x^\mu\}$. In $D=(1+d)$ dimensions this is given by
\be
f_t=\sqrt{|\mathsf{g}_{tt}|}=\int d\xi^d\sqrt{|\mathsf{g}|}e^{-\pi\mathsf{g}_{ij}\xi^i \xi^j}\,, \quad \mathsf{g}_{ti}=0\,,
\ee
in a co-moving reference frame.

\subsubsection*{Momentum distributions}
A consequence of that the particle's position relative to a reference frame $\{x^\mu\}$ follows the probability $P(x^\mu)$ in \eqref{eq.Ppos} is that the momentum of the particle is set by Heisenberg's uncertainty principle to obey a probability distribution of
\be\label{eq.Pmom}
P(P^\mu)=(\pi\hbar)^{-D}\sqrt{|\mathsf{g}^{-1}|}e^{-\mathsf{g}^{-1}_{\mu\nu}\zeta^\mu\zeta^\nu/(\pi\hbar^2)}\bigg|_{\zeta^u\rightarrow i \zeta^u}\,,\,\,\,\, \zeta^\mu=P^\mu-P^\mu_o\,, \,\,\,\, \mathsf{g}_{\mu\nu}^{-1}=\mathsf{g}^{\rho\sigma}\delta_{\rho\mu}\delta_{\sigma\nu}\,,
\ee
relative to the same co-moving reference frame $\{x^\mu\}$, where $P^\mu_o$ is the average momentum of the particle, relative to the reference frame. The reason for this is as follows. Given a particle position distribution as in \eqref{eq.Ppos}, the particle's momentum distribution is the minimal distribution allowed by Heisenberg's uncertainty principle (no extra uncertainty added) in the presence of said position distribution. The minimality condition sets the shape of the momentum distribution to also be a Gaussian, and for a general Gaussian distribution, Heisenberg's uncertainty principle constrains a particle with its position given by the $P(x^\mu)$ in \eqref{eq.Ppos} to be characterised by a $P(P^\mu)$ given by
\be
\pi^{-D}\sqrt{|\mathsf{h}|}e^{-\mathsf{h}_{\mu\nu}\zeta^\mu\zeta^\nu/\pi}\bigg|_{\zeta^u\rightarrow i \zeta^u}\,, \quad \mathsf{h}_{u\mu}\stackrel{u\neq\mu}{=}0\,, 
\ee
\be\label{eq.Hun}
\mathsf{g}_{\mu\nu}\,,\,\mathsf{h}_{\rho\sigma}\,:\quad {\rm tr}(\mathsf{g}_{\mu\nu}\delta^{\nu\rho}\mathsf{h}_{\rho\sigma})\leq D\hbar^{-2}\,,\quad \mathsf{gh}\leq \hbar^{-2D}\,.
\ee
Here, the uncertainty in momentum is minimised by $\mathsf{h}_{\mu\nu}=\mathsf{g}_{\mu\nu}^{-1}/\hbar^2$. This defines what the particle is limited by at the quantum level, and hence it also defines what the particle is characterised by the at the quantum level, where the only uncertainty present is due to the quantum properties of the particle.

That Heisenberg's uncertainty principle\footnote{Note that the $\sigma_t$ here refers to the standard deviation for 
where an interaction takes place in time.}
\be
\sigma_x \sigma_{p_x}\geq \hbar/2\,,\quad \sigma_t\sigma_E\geq \hbar/2
\ee
corresponds to \eqref{eq.Hun} can be seen as follows. The multi-dimensional Gaussian functions used here constitute an extension of the normalised, single variable Gaussian function
\be
\frac{1}{\sigma_x\sqrt{2\pi}}e^{-\frac{(x-\mu)^2}{2\sigma_x^2}}\,,
\ee
so the entries of a diagonal $\pi\mathsf{g}_{\mu\nu}$ corresponds to $(2\sigma_{x^\mu}^2)^{-1}$, and Heisenberg's uncertainty principle equivalently is
\be\label{eq.exH}
(2\sigma_{x^\mu}\sigma_{P^\nu})^{-2}\leq \hbar^{-2}\,, \quad\forall\,\mu=\nu\,.
\ee
Moreover, the $(\mathsf{g}_{\mu\nu},\mathsf{h}_{\mu\nu})$ are diagonal in the same reference frames. This is most straightforward in the reference frame $\{\tilde x^\mu\}$, where $\mathsf{g}_{\mu\nu}=\eta_{\mu\nu}$ and $\mathsf{h}_{\mu\nu}$ is diagonal, with $\mathsf{g}_{\mu\nu} \mathsf{h}_{\rho\sigma}=\eta_{\mu\nu} \mathsf{h}_{\rho\sigma}\leq \hbar^{-2}$ for $\mu=\nu=\rho=\sigma$. Any other reference frame is just a transformation of that local, co-moving frame. If we treat $\mathsf{g}_{\mu\nu}$ and $\mathsf{h}_{\mu\nu}$ as matrices, Heisenberg's uncertainty principle concerns the eigenvalues of those matrices, and since the matrices are simultaneously diagonalisable, it then follows that \eqref{eq.Hun} is equivalent to \eqref{eq.exH}.
\\\\
In total, the $P(x^\mu)$ and $P(P^\mu)$ of \eqref{eq.Ppos} and \eqref{eq.Pmom} provide a setting consistent with the conjecture in \cite{Karlsson:2019vpr}, on how quantum uncertainties for the spacetime position and momentum of a particle might be modelled. Note that the $\mathsf{g}_{\mu\nu}$ allowed by different reference frames through \eqref{eq.refch} includes a free scale in each $\sigma_{x^\mu}$ which is irrelevant to the physics, since changes encompassed by \eqref{eq.refch} purely refer to how the physics appears in relation to the reference frame set by $\{x^\mu\}$. The physically relevant properties at the quantum level are internal: relative within the spacetime, and insensitive to the choice of $\{x^\mu\}$.

\section{How the quantum model gives a metric and the geodesic equation}\label{s.rise}
Given the probability distributions that the quantum fluctuations are characterised by in a reference frame $\{x^\mu\}$, it is straightforward to identify that they can be used as basis functions for the spacetime. We discuss this below, before we describe how the random walk of the quantum fluctuations defines their propagation through spacetime, and how that propagation turns out to be specified by the geodesic equation.

\subsection{The metric}\label{s.met}
The equation \eqref{eq.locr} explicitly shows how each quantum fluctuation defines a local concept of spacetime. In interactions between particles, it is the Gaussian probability distributions in spacetime that mediate this concept of spacetime --- any particle in the spacetime moves relative to the average behaviour of the quantum fluctuations in its vicinity. This means that the probability distributions effectively play the role of basis functions, through
\be\label{eq.met}
g_{\mu\nu}(x^\rho)=\langle \mathsf{g}_{\mu\nu}\rangle
\ee
where the average is over a small spacetime volume element around $x^\rho$. The spacetime can also be visualised as given by the surface specified by the basis functions on $\{x^\mu\}$, same as e.g. any smooth function $f(x)$ can be specified by Gaussian basis functions; only for the spacetime surface, all the basis functions have the same weight.

The line element is given by
\be\label{eq.ds2}
ds^2=g_{\mu\nu}dx^\mu dx^\nu
\ee
and is characterised by that it overlaps with \eqref{eq.locr} at any given point. The metric has an inverse, and for any point $x_o^\rho$ in spacetime a reference frame can be found where
\be
ds^2=\eta_{\mu\nu}dx^\mu dx^\mu+\mathcal{O}(\epsilon)\,,\qquad \epsilon^2=\eta_{\mu\nu}(x^\mu-x_o^\mu)(x^\nu-x_o^\nu)\,.
\ee

The above shows how the GR line element, with its specific properties, can arise from interactions between quantum fluctuations, and an invariance of the encoded interactions between the particles, as observed for \eqref{eq.refch}. In this setting, the diffeomorphisms of GR represent the different reference frames that provide equivalent descriptions of the internal dynamics – the dynamics that is present in-between the quantum fluctuations. At the quantum level, the only physical processes are relative; set in relation to the behaviour of the particles present in the spacetime. The reference frame itself is irrelevant, as long as it captures the correct internal processes, i.e. \eqref{eq.locr}. In fact, the gauge/gravity duality is in line with that this insensitivity to the chosen reference frame even extends to models where the number of dimensions in $\{x^\mu\}$ has been reduced; as long as the internal physics is captured, any reference frame can be used. In the present text, we however restrict the discussion to full $D$-dimensional reference frames. Here, we also have that the movement of a spacetime volume element relative to the reference frame $\{x^\mu\}$ is set by the momentum of a spacetime volume element, $ \langle P_o^\mu\rangle$, where the average is over a small spacetime volume element (same as for the metric).

From \eqref{eq.met} it is also clear that the metric is not well-defined at very small scales, which indeed is an integral feature of the present ansatz for spacetime emergence. When the scale under consideration approaches the scale of the probability distributions, quantum effects will appear in different forms. A first sign of this will be that individual distributions (and their statistical deviations) have more of an impact on the $ds^2$. This in turn initiates a transitional phase to the scale of the actual individual distributions, where the spacetime concept is not well-defined, and the physics instead is governed by interactions between the quantum fluctuations. As argued in \S\ref{s.Gprob}, there are reasons to expect the standard deviations of the Gaussian distributions to be of order the Planck scale.

Note that the discussion above concerns regions of spacetime where only quantum fluctuations are present\footnote{This is not quite the same as vacuum regions of spacetime, which are defined by $R_{\mu\nu}=0$.}. In the present text, this is the only type of spacetime region that we consider. Anything out of the vacuum is modelled through a boundary condition, as described in \S\ref{s.Ricci}. It is however clear from the set-up that individual particles (e.g. photons) that travel through the spacetime will interact with the quantum fluctuations; this is how particles passing through will experience and be subject to the configuration of the spacetime. At interaction with the quantum fluctuations, those passing particles will also have an impact on the behaviour of the quantum fluctuations, and the metric will change because of it.

\subsection{The random walk of the quantum fluctuations}\label{s.rWalk}
At each new, initiated interaction, the propagation of a quantum fluctuation in spacetime is described by a Gaussian random walk. The step length is set by the $P_o^\mu$ of the particle, and the Gaussian probability distribution in \eqref{eq.p1}. However, relative to a general reference frame $\{x^\mu\}$ the random walk is biased, e.g. since the $\sigma$ of the Gaussian is set relative to the incoming interaction, from the environment. On average, the random walk of a quantum fluctuation is specified by the metric, $g_{\mu\nu}$.

In the below, we take a close look at what the random walk of the quantum fluctuations is characterised by, before connecting the behaviour to the geodesic equation. Note that a key feature of the random walks of the quantum fluctuations is that they are Gaussian random walks relative to the internal dynamics, which defines what is physical at the quantum level. In setting the spacetime propagation of a quantum fluctuation relative to the internal dynamics, we can use the general properties that Gaussian random walks are characterised by.

\subsubsection*{Gaussian random walks}
Some of the standard concepts of Gaussian random walks have already been described in \S\ref{s.eqmod}. In the quantum model, each step is given by the sum of a mean contribution $(\propto P_o^\mu)$ and a random contribution generated by the Gaussian distribution in \eqref{eq.p1}, which is set relative to the local rest frame of the particle (the $S^{D-1}$ frame of the particle). 

The standard example of a Gaussian random walk in one dimension is that each step is generated by the probability distribution
\be\label{eq.G1}
\frac{1}{\sigma_x\sqrt{2\pi}} e^{-\frac{(x-\mu)^2}{2\sigma_x^2}}\,.
\ee
where $\mu$ is the mean step length and $\sigma_x$ is the standard deviation. After $n$ steps have been generated in this manner, a particle that initially was located at $x=0$ will have a position given by
\be
P(x)=\frac{1}{\sigma\sqrt{2\pi}} e^{-\frac{(x-n\mu)^2}{2\sigma^2}}\,,\quad \sigma^2=n\sigma_x^2\,.
\ee
This illustrates how the probability distribution for the position of the particle is given by a Gaussian distribution that spreads out during the propagation of the particle, when each step is given by a Gaussian distribution. In the below, we will use both of these properties: that the probability distribution for the random walk effectively specifies both the individual steps in the random walk, and the result after a series of steps.

\subsubsection*{The biased random walk}
As described in \S\ref{s.qmod}, the quantum fluctuations are represented relative to a $D$-dimensional reference frame $\{x^\mu\}$, under the assumption that the interactions are sufficient to correlate the quantum fluctuations. The only requirement on the $\{x^\mu\}$ is that it captures the internal, relative dynamics of the quantum fluctuations, and leaves that internal dynamics unaltered; $ds^2$ must be unchanged, as in \eqref{eq.refch} and \eqref{eq.ds2}. As a consequence, the position, density and interaction frequency of the quantum fluctuations relative to $\{x^\mu\}$ depends on the choice of $\{x^\mu\}$. 

For example, the average interaction rate at a point $x^\rho$ in spacetime is 
\be\label{eq.ftdef}
f_t\;:\quad g_{tt}(x^\rho)=-f_t^2(x^\rho)
\ee
where $t$ is a time-like variable, and we recall that $g_{\mu\nu}(x^\rho)$ is an average of the $\mathsf{g}_{\mu\nu}$ in a spacetime volume element around $x^\rho$. This function of $g_{tt}$ as the square of a frequency directly comes from that $g_{tt}dt^2$ is a (scale-invariant) contribution to $ds^2$. Depending on the choice of $\{x^\mu\}$, $f_t$ can vary both from a change in the density of quantum fluctuations relative to $\{x^\mu\}$, and due to that the individual interaction rates of the quantum fluctuations change with $x^\mu$.

A key feature of the Gaussian random walk of a quantum fluctuation is that the properties of the random walk are set relative to the $S^{D-1}$ of the effective quantum model, at each step. This means that when the interaction rate $f_t$ of the quantum fluctuations varies relative to the reference frame, $f_t=f_t(x^\mu)$, the random walk as seen from the reference frame will be biased, even when the particle is not subject to external forces (e.g. an electric field). 

What happens when the random walk is described relative to a reference frame $\{x^\mu\}$ is that each single quantum fluctuation will appear to move through a medium with varying interaction density (and length scales), and since its step length (and direction) is set relative to a particle-specific unit of incoming interaction (not relative to $\{x^\mu\}$), the step length and direction will be distorted relative to $\{x^\mu\}$. The density can be pictured in terms of the Gaussian probability distributions. In their role as basis functions, the distributions are centred at a point $x_o^\rho$ in spacetime, and extend away from it. The distributions overlap, and provide an effective `density' of interactions. From the perspective of $\{x^\mu\}$, the effective function is given by $g_{\mu\nu}(x^\rho)$.

\begin{figure}[tbp]
\includegraphics[width=0.47\textwidth]{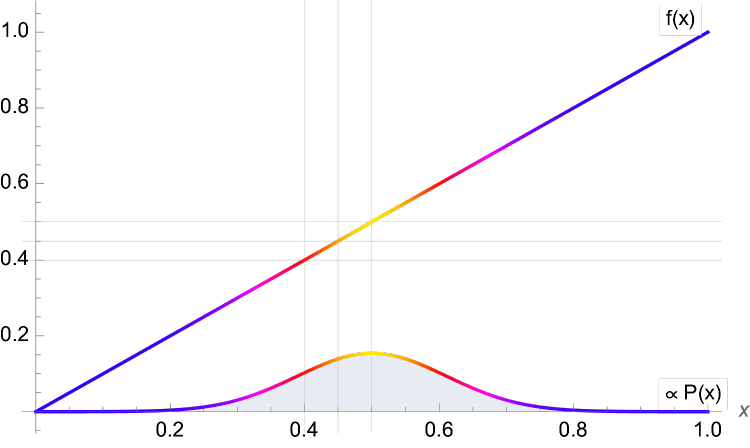}
\hspace{0.04\textwidth}
\includegraphics[width=0.47\textwidth]{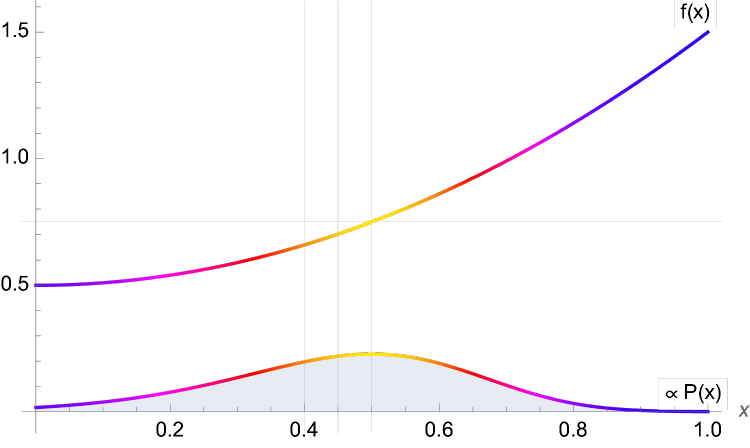}
\caption{In each graph we see a surface $f(x)$ with a Gaussian distribution on it (coloration). To the left, the surface has $\partial_x^2 f(x)=0$ and follows the equation $f(x)=x$. The distribution is centred around $x=0.5$ and has standard deviation $\sigma=1/3$. To the right, the surface follows the equation $f(x)=x^2+0.5$, and the distribution has $\sigma=1/2$. Below each surface is an illustration of the distribution as seen from the $x$-axis (for clarity, these are not normalised). To the left, the Gaussian distribution on the surface is a Gaussian distribution also relative to the $x$-axis, but to the right the distribution is distorted. It is clearly visible that the Gaussian distribution on the bent curve translates into a non-Gaussian function relative to the $x$-axis. In the graph to the right, the mean value of $x$ is no longer $0.5$. Instead, $\langle x\rangle<0.5$. In addition, with an increase in $\sigma$ over time ($\partial_t\sigma>0$), the graph to the right is characterised by $\partial_t \langle x\rangle<0$, so that $\langle x\rangle$ moves to lower values. In the same scenario, the graph to the left has $\langle x\rangle=0.5$, $\partial_t \langle x\rangle=0$. What we want to illustrate with this picture is that a particle that moves on the surface according to a Gaussian random walk will have a position $\langle x(t)\rangle$ that depends on the shape of the surface. In this $1d$ example, $\partial_x^2 f\neq0$ causes a shift in $\langle x(t)\rangle$. In general, it is $\partial_\rho g_{\mu\nu}\neq0$ that causes this type of effect. For example, the moving particle can be a quantum fluctuation that starts out at $x=0.5$ with a Gaussian probability distribution for its position that is characterised by a $\sigma_x$ at a scale small enough for $\partial_x f$ to be well approximated by a constant. Relative to $\{x^\mu\}$, the particle will be subject to an interaction rate $f_t=\partial_x f$ from its environment. After $n$ steps in a Gaussian random walk with variance $\sigma_x^2$, the variance of the position distribution will be $\sigma^2=n\sigma_x^2$, and in that sense the $\langle x\rangle$ will evolve in time. Note that the evolution of $\langle x\rangle$ only takes place relative to $\{x^\mu\}$. The centre point of the distribution as seen from the surface $f(x)$ remains constant (at $x=0.5$ in the graphs).\label{fig.distort}}
\end{figure}

For a detailed illustration of the biased random walk, it is useful to consider a scenario with a series of random steps, where the interaction frequency only depends on one spatial dimension, $f_t=f_t(x)$. We also disregard contributions from $P_o^\mu$, equivalently the $\mu$ in \eqref{eq.G1}. The setup is depicted in figure \ref{fig.distort}. 
The random walk can effectively be viewed as taking place on a surface, $f(x)$, with the interaction rate given by $f_t=\partial_x f(x)$. When $f_t$ is a constant ($\partial_xf_t=0$) standard Brownian motion applies. The interaction frequency (and hence the step length of the random walk) relative to $x$ can be altered through a rescaling of $x$ without any physical effect. However, when $\partial_xf_t\neq 0$ the mean value of the position of the quantum fluctuation relative to $x$,\footnote{Here, the probability distribution $P_q$ is initially given by the $1d$, coordinate transformed version of \eqref{eq.p1}. As the random walk progresses, the $\mathsf{g}_{\mu\nu}$ transitions into $g_{\mu\nu}$. The variance also scales with $(1+n)$ after $n$ steps.}
\be
x_q=\langle x\rangle_{P_q}=\int dx\, x P_q(x)\,, 
\ee 
will shift with time --- in this example, time is present in terms of the number of random steps the quantum fluctuation has taken. When the Gaussian spreads on a surface with $\partial_x f_t\neq0$, the expected position of the particle (in the reference frame) changes.

The propagation of the expected position of a quantum fluctuation that is driven by the random walk is present at each step of the random walk, and it is valid for all of the spacetime directions. Even time propagation can be depicted as in figure \ref{fig.distort}, with the probability distribution in time obtained in the same way as in \S\ref{s.Gprob}. $x_q^\rho$ simply denotes the expected point in spacetime, modulo the translation(s) set by $P_o^\mu$, at which a quantum fluctuation initiates an interaction with its surroundings. An alteration of the $x_q^\rho$ at a random step means that the quantum fluctuation accelerates relative to $\{x^\mu\}$. This movement has one of two origins; either it is induced by the choice of reference frame, or due to non-trivial physics. The analogy in GR is apparent: the acceleration must be in one-to-one correspondence with the geodesic equation.

\subsection{The geodesic equation}\label{s.geo}
Given a metric, as identified in \eqref{eq.met}, the geodesic equation 
\be\label{eq.geod}
\frac{d^2x^\mu}{ds^2}+\Gamma^\mu_{\rho\sigma}\frac{dx^\rho}{ds}\frac{dx^\sigma}{ds}=0
\ee
can be derived through standard procedures (equivalence principle, parallel transport etc.). Below, we detail how the random walks of the quantum fluctuations give rise to an acceleration of the expected position of each particle that is captured by the geodesic equation, and we give a detailed calculation for the acceleration caused by $P^t_o\neq0$, equivalently $dt\neq0$. This illustrates the principles involved for $dx^\mu\neq0$ very well --- the identifications follow from the same procedures --- so the contribution from $dt\neq0$ is sufficient to illustrate how the geodesic equation arises from the perspective of the random walks of the particles in the spacetime. As always, the acceleration contributions add up to give the total acceleration.
\\\\
To begin with, note that the momentum and spacetime position of any particle present in the spacetime, be it a quantum fluctuation or an `external' particle, are described by probability distributions due to Heisenberg's uncertainty principle (as illustrated for quantum fluctuations in \S\ref{s.Gprob}). At the quantum level, the step length and direction of each regular displacement $dx^\mu$ of the particle (set by $P_o^\mu$) varies a bit. Classical notions do not deal with these quantum level probability distributions. Instead, the effective classical notions of position, momentum and displacements $dx^\mu$ refer to averages: the expectation values set by the quantum probability distributions. The spacetime probability distribution of a particle makes its propagation sensitive to $\partial_\rho g_{\mu\nu}\neq0$, which causes shifts in the expected position of the particle. In general, it is $g_{\mu\nu}$ that encode the spacetime scales, and take the role of scale that the interaction rate $f_t: f_t^2=-g_{tt}$ played in the example in figure \ref{fig.distort}; the principles identified are the same. While we focus on the propagation of the quantum fluctuations in the present text, the concepts apply more generally, to particles moving through the spacetime.

We have already illustrated what happens with the expectation value of the position of a quantum fluctuation under time evolution in \S\ref{s.rWalk} and in figure \ref{fig.distort}. It is the same process that from a random walk perspective gives rise to the geodesic equation. In general, one can picture the spacetime as a surface relative to a Minkowski reference frame $\{x^\mu\}$ (but in a higher dimension). On that surface, the particles move in a Gaussian random walk, and when the surface bends relative to $\{x^\mu\}$, the particles get accelerated, relative to $\{x^\mu\}$. In a general reference frame, a change in $x_q^\mu$ due to a gradient in the interaction frequency and notion of length relative to $\{x^\mu\}$, i.e. $\partial_\rho g_{\mu\nu}\neq0$, will change the $dx^\mu$ for the particle, giving $d^2 x^\mu/ds^2\neq0$. For an analysis of $d^2 x^\mu/ds^2$, it is sufficient to restrict to the effect $\partial_\rho g_{\mu\nu}\neq0$ has on the end point, at a random step. The acceleration caused by the random walk in this way can be shown to correspond to the geodesic equation. Below, we detail how $dt\neq0$ contributes to the acceleration.

In general, note that in the present model the movement of any particle in spacetime is decided purely relative to the internal dynamics of the spacetime, i.e. relative to the surrounding quantum fluctuations and their interaction profiles; the local perception of spacetime is set relative to the $S^{D-1}$ frames of the quantum fluctuations. The reference frame employed does not affect the local dynamics, only the depiction of it. At the quantum level, the physical entities are effectively defined relative to the internal dynamics, not relative to the choice of reference frame. It is at the classical level that effects relative to the reference frame, such as gravitational acceleration of massive, classical objects, become relevant.

\subsection*{Illustration: acceleration induced by $dt$}
We will now detail how the random walk of a particle infers an acceleration of its expected position that is described by the geodesic equation, for the special case where only $dt\neq0$. For this illustration, we restrict the spacetime to $D=(1+d)$ dimensions.

To begin with, note that the acceleration $d^2 x^\mu/ds^2$ induced by $dt\neq0$ is caused by two separate processes: {\it (i)} the Gaussian random walk of the particle, and {\it (ii)} the movement of rest frame of the particle relative to the reference frame.
\\\\
For the spatial acceleration, $d^2 x^i/ds^2$, the first effect was described and depicted in and around figure \ref{fig.distort}. In the reference frame where $g_{\mu\nu}(x_o^\rho)=\eta_{\mu\nu}$ and the normalisations are simple, that contribution is 
\be\label{eq.peq}
-\partial_i f_{t}\bigg|_{x^\rho=x_o^\rho}
\ee
since the local acceleration of the expected position is given by $\partial_i f_t$ (which can be thought of as a gradient in pressure), and a higher interaction frequency towards one side means an acceleration in the opposite direction.

In addition, the rest frame of the particle can move with a velocity $v^i$ relative to the reference frame. This gives both the direct, straightforward acceleration contribution of $\partial_t v^i$, as well as a contribution of $v^i \partial_t f_t$ when the interaction rate of a particle changes in time $(\partial_t f_t\neq0)$. The last contribution comes from that a decrease in interaction rate ($\partial_t f_t<0$) corresponds to a de-acceleration in the direction of the velocity, since the points at which the particle interacts will grow sparser relative to $\{x^i\}$ as time progresses. That provides a separate process for how $\langle x^i\rangle$ is altered. Note that the last two contributions do not arise because the velocity or the interaction rate of the particle changes, but because those properties are altered relative to the reference frame due to that the reference frame is not flat.
In total, 
\begin{subequations}
\label{eq.xoD}
\be
\label{eq.xoda}
\left(\frac{dt}{ds}\right)^{-2}\frac{d^2 x^i}{ds^2}\bigg|_{x^\rho=x_o^\rho\,,\,dx^j=0}=\left(-\delta^{ij}\partial_j f_t+\partial_t v^i+ v^i\partial_t f_t\right)\bigg|_{x^\rho=x_o^\rho}\,.
\ee
This relation can be translated into a general equation through identifying that the metric encodes the relevant entities through $-g_{tt}=f_t^2$ and $g_{ti}=- v_i$. The first of these identifications is the same as in \eqref{eq.ftdef}. The second identification can be deduced from the simple set-up where the rest frame of the particle, $\{\tilde x^\mu\}$, moves relative to the reference frame $\{x^\mu\}$ with $x^i=\tilde x^i+w^i(x^\rho)$, where $v^i=\partial_t w^i$, and with $t=\tilde t/\sqrt{1-\eta_{ij}v^i(x_o^\rho) v^j(x_o^\rho)}$ and $\partial_j w^i(x_o^\rho)=0$. In that setting, $ds^2=\eta_{\mu\nu}d\tilde x^\mu d\tilde x^\nu$ sets $g_{ti}=-v_i$ and leaves the other components of the metric unchanged at $x_o^\rho$. In general settings, the principle remains the same: the $v_i$ present in \eqref{eq.xoda} corresponds to $-g_{ti}$. Using these two identifications as well as the special choice of reference frame, the right hand side of \eqref{eq.xoda} can be identified to equivalently be
\be\label{eq.xodb}
\left[-\delta^{ij}\frac{1}{2}\partial_j (-g_{tt})+\delta^{ij}\partial_t(-g_{tj})+\delta^{ij}g_{tj}\eta^{tt}\frac{1}{2}\partial_t(-g_{tt})\right]\bigg|_{x^\rho=x_{o}^\rho}\,.
\ee
\end{subequations}
Here, the third term has an index structure that captures the $D=(1+d)$ illustration at hand, but leaves out details necessary for general spacetimes. For the same reason, the index notation looks messy with respect to $t$, but the expression is valid since $t$ just represents the temporal dimension.

From \eqref{eq.xoD}, the general equation can be identified to be
\be
\frac{d^2 x^i}{ds^2}\bigg|_{dx^j=0}=\left[\frac{1}{2}g^{ij}\partial_j g_{tt}-g^{ij}\partial_t g_{tj}-\frac{1}{2}g^{it}\partial_t g_{tt}\right]\left(\frac{dt}{ds}\right)^2=-\Gamma^i_{tt}\left(\frac{dt}{ds}\right)^2\,,
\ee
for a completely general metric $g_{\mu\nu}$. This follows from that \eqref{eq.xodb} holds for all $x_o^\rho$ provided that the metric has been altered to $g_{\mu\nu}(x_o^\rho)=\eta_{\mu\nu}$, modulo a velocity. Including extended regions, as in an expansion around $x^\rho=x_o^\rho$, alters the $\eta_{\mu\nu}$ components to $g_{\mu\nu}$, and that results in the general expression.
\\\\
The temporal acceleration $d^2t/ds^2$ can be identified in a similar way. The components of $x^\mu_q$ (expected spacetime position modulo changes by $P_o^\mu$) are altered by the same processes: the probability distribution extends over a region in $\{x^\mu\}$, and $\partial_\rho g_{\mu\nu}\neq0$ distorts that probability distribution relative to $\{x^\mu\}$. 

For simplicity, we follow the same line of reasoning as for the spatial acceleration described above, and use the same type of reference frame. There are three different contributions to the temporal acceleration induced by $dt\neq0$. The first two contributions are due to the change in $f_t$ in the region the particle moves into, $-\partial_t f_t -v^i \partial_i f_t$. The first of these is the counterpart of \eqref{eq.peq} in our previous example. The second term comes from that with a non-zero $v^i$, $dt$ causes a transport that is not only in time, but has a spatial component as well. In comparison with the illustration in figure \ref{fig.distort}, in $2D$ the surface there would need to be extended to depend on both dimensions, $f=f(t,x)$, and $dt\neq0$ would cause a transport $(t_i,x_i)\rightarrow (t_f,x_f)$ with the line of propagation set by $v^i$. Under that type of displacement, the gradient in $f_t$ that causes a shift in $t_q$ is given by\footnote{Note that the contribution of $(-\partial^i+v^i\partial_t) f_t$ in \eqref{eq.xoda} is of the same type. The acceleration contributions can also be inferred/interpreted from other perspectives than that of the random walk.} $-\partial_t f_t -v^i \partial_i f_t$. Finally, a change in time dilation also gives a change in interaction rate relative to $\{x^\mu\}$. Since $f_t^2$ includes the term $-v^i v_ i$ (as in $f_t^2=f_t^2\big|_{v^i=0}-v^i v_i$) that contribution is $v^i\partial_t v_i$. In total,
\begin{subequations}
\begin{align}
\left(\frac{dt}{ds}\right)^{-2}\frac{d^2 t}{ds^2}\bigg|_{x^\rho=x_o^\rho\,,\,dx^j=0}&=\left(-\partial_t f_t-v^i \partial_i f_t+v^i\partial_t v_i\right)\bigg|_{x^\rho=x_o^\rho}\\
&=\left(-\frac{1}{2}\eta^{tt}\partial_t g_{tt}+\frac{1}{2}\eta^{tt}g_{tj} \delta^{ji}\partial_i g_{tt}-\eta^{tt}g_{tj}\delta^{ji}\partial_t g_{ti}\right)\bigg|_{x^\rho=x_o^\rho}\,,
\end{align}
\end{subequations}
from which the general equation can be identified,
\be
\frac{d^2 x^t}{ds^2}\bigg|_{dx^j=0}=\left[-\frac{1}{2}g^{tt}\partial_t g_{tt}+\frac{1}{2}g^{ti} \partial_i g_{tt}-g^{ti}\partial_t g_{ti}\right]\left(\frac{dt}{ds}\right)^2=-\Gamma^t_{tt}\left(\frac{dt}{ds}\right)^2\,.
\ee

\section{Ricci-flat spacetimes}\label{s.Ricci}
The crucial property for any ansatz for spacetime emergence is that the effective theory, the theory that emerges at large scales, must be GR. As our ansatz is for how quantum fluctuations give rise to spacetime, the emergent effective theory must capture GR in vacuum regions, characterised by 
\be
R_{\mu\nu}=0\,.
\ee
A caveat is of course that there might be circumstances under which the effective GR theory does not (and should not) emerge, i.e. when the GR theory breaks down. Here, we however work under the assumption that a metric has materialised. It it then possible to show that the effective quantum model gives a spacetime with $R_{\mu\nu}=0$, as described in \S\ref{s.qRicci} below.

Importantly, a vanishing Ricci tensor is only half of the story of how a specific spacetime solution arises in a vacuum region. $R_{\mu\nu}=0$ is a local concept. The overall characteristics of a spacetime solution are in turn set by the configuration of the non-vacuum features present in the spacetime. In our effective model, the presence of matter --- and in general any feature that is not captured by quantum fluctuations in a region of spacetime --- is modelled by boundary conditions to the vacuum region under consideration. These boundary conditions are of the sort used for solutions to the heat equation, and literally represent boundaries to the vacuum regions. We describe the impact of the boundary conditions on the spacetime configurations of the vacuum regions in \S\ref{s.bound}. At the current level of our ansatz, the boundary conditions are effective only, i.e. they are characterised by how they effectively influence the adjacent vacuum regions. This is sufficient for the purposes of the present text, but of course, it would also be interesting to understand the related interaction processes in detail, at the quantum level.

\subsection{How the effective quantum model comes with a vanishing Ricci tensor}\label{s.qRicci}
The condition that the Ricci tensor vanishes shows as a product of the random walk of the quantum fluctuations when the internal particle propagation follows the effective quantum model in \S\ref{s.eqmod}. It requires an absence of separate processes for the particle propagation, such as are due to additional forces. The reason is as follows. The Gaussian random walk means that a particle with a momentum $P_o^\nu$ does not propagate solely in the direction of the corresponding velocity, $v^\nu$. Any step $dx^\nu$ in the direction of $P^\nu_o$ comes with another randomly generated step. The order of the two steps have equal probability. This generates an effective result
\be\label{eq.pProp}
(v_f)_\mu=\frac{1}{2}\left( D_\mu D_\nu+D_\nu D_\mu\right)v^\nu\,,
\ee
where $v_f^\mu$ is the final vector, $v^\nu$ is the vector subject to transport, the parallel transport along a vector is captured by $D_\mu$, and the factor $1/2$ denotes the probability for each order of the parallel transports. This holds for all $\mu$, since the random propagation covers every possible additional step. Moreover, the relation holds for all $v^\nu$, where each quantum fluctuation has an individual $P^\nu_o$ with a statistical deviation from the $\langle P_o^\nu\rangle$ which characterises a small spacetime volume element. The random walk of the quantum fluctuations forces \eqref{eq.pProp} to be a property of the spacetime for all the individual $P_o^\nu$, and for all possible random steps $dx^\mu$. Since the end product of a displacement of $v^\nu$ is symmetric with respect to the order of parallel transport, the antisymmetric counterpart vanishes,
\be\label{eq.comm}
[D_\nu,D_\mu]v^\nu=0\,,\quad \forall v^\nu\,,
\ee
and since
\be
[D_\nu,D_\mu]v^\nu=R_{\mu\nu}v^\nu\quad \text{if}\quad v^\nu\neq0\,,\,\forall \nu\,,
\ee
in combination with that each spacetime volume element contains a statistical variation in $v^\nu$,
\be
\eqref{eq.comm}\quad\Rightarrow \qquad R_{\mu\nu}=0\,.
\ee
Recall that the quantum fluctuations, and their random walks, generate the spacetime. Hence the spacetime is defined by $R_{\mu\nu}=0$.

In contrast, if a separate process for particle propagation is introduced, \eqref{eq.pProp} no longer is a requirement, by that it is not by necessity enforced by the physical processes present in the spacetime. One such scenario, discussed a bit more in the appendix, is when charged quantum fluctuations are subject to a force in the presence of an electric field. In that case, a separate flow of particles is introduced.

Moreover, a condition for that the Riemann tensor vanishes is a special case of $R_{\mu\nu}=0$, where
\be
[D_\mu,D_\nu]u^\rho=R^\rho{}_{\sigma\mu\nu}u^\sigma=0\,,\quad \forall u^\rho\,,
\ee
and it differs from the $R_{\mu\nu}=0$ condition observed above in that parallel transport of any vector $u^\rho$, in any two directions, commutes. One of the directions for the parallel transport does not need to be parallel to the vector. In that sense, the statement that the two displacements commute (insensitivity to order of displacement) is not connected to particle propagation; it is a stricter condition. It characterises certain vacuum regions, but its presence is contingent upon a presence of the right boundary conditions.

\subsection{Boundary conditions}\label{s.bound}
The quantum fluctuations interact with each other in the same way in all spacetime regions with $R_{\mu\nu}=0$. Different spacetime solutions arise from what the boundaries of the vacuum regions are characterised by, relative to the reference frame.

The simplest type of spacetime is flat spacetime, where the Riemann tensor vanishes. It arises when the presence of matter at the boundaries can be neglected. Either the boundary is infinitely far away (the vacuum spacetime region extends to infinity) or the impact of the matter at the boundary can be neglected at the scales under consideration (when the curvature only is relevant closer to the boundary than the region under consideration). However, a vanishing Riemann tensor does not mean that the spacetime configuration is trivial. As discussed in \S\ref{s.mot}, in the presence of boundaries restricting information exchange, the present ansatz for spacetime emergence comes with non-trivial configurations of flat spacetime, set relative to how said boundaries extend in the spacetime \cite{Karlsson:2019avt}. In the present text, we do not discuss those configurations though. Instead, we focus on the key boundary features that give different types of solutions with $R_{\mu\nu}=0$.

Note that the boundary conditions we describe represent a simplified model for how the boundaries and the quantum fluctuations interact, on average and at an effective level. This is sufficient for the purpose of the present text which is to show that, and how, spacetime can emerge from interactions between quantum fluctuations. However, for a full picture of the physics it would be desirable to have more detail on how the quantum fluctuations interact with particles (and extended objects) that are not part of the vacuum regions. 

In the current model, anything present in the spacetime that is not a quantum fluctuation generated by the vacuum calls for a boundary condition. In the below, we use the simplest possible settings to illustrate the relevant properties. In principle, a combination of these for multiple boundaries of various relative positions and configurations (including the shape of a boundary) can be used to capture general spacetime settings. However, the actual procedure for doing so, and for determining the metric of the resulting spacetime, is highly non-trivial.

In setting up a set of boundary conditions to a vacuum region, and to illustrate the role of different types of boundary conditions, the simplest type of example is to start from a spacetime that is asymptotically flat. In that configuration, the vacuum region extends very far away and all non-vacuum features are centred in a comparatively small region of the spacetime, near the centre of the reference frame. In this setting, we illustrate the two types of boundary conditions that are key to spacetime regions with $R_{\mu\nu}=0$: mass and relative movement of the boundaries. We also use $D=(1+3d)$ for the explicit examples; that is sufficient for an illustration of the general principles involved.

\subsubsection*{Massive vs not: a Neumann boundary condition}
The simplest boundary condition to add to the initial set-up is a $S^{d-1}$ sphere that is static in time, at rest and located at the centre of the reference frame. In our ansatz, the presence of mass in the region bounded by the sphere is reflected by that the boundary condition is characterised by
\be\label{eq.condm}
dx^t dx^a \partial_a f_t=const>0\,
\ee
where $dx^a$ is a vector perpendicular to the surface given by the boundary, pointing into the spacetime region. This Neumann boundary condition imposes a gradient in the interaction frequency $f_t$ of the spacetime, which is absent in flat spacetimes, and causes curvature. In terms of the metric, $g_{tt}=-f_t^2$, and the Neumann boundary condition rules out $\partial_\rho g_{\mu\nu}=0$ as an option for any reference frame --- except as an approximate solution, far away from the boundary. The metric therefore is not characterised by $ds^2=\eta_{\mu\nu}dx^\mu dx^\nu$, and the Riemann tensor is non-zero. For a flat spacetime, it is necessary that all boundary conditions are characterised by
\be
\partial_a f_t=0\,.
\ee
Moreover, the condition in \eqref{eq.condm} causes an attractive force towards the surface, relative to the reference frame, since the random walks of the quantum fluctuations are biased towards lower interaction frequency; the geodesic equation will describe acceleration towards the boundary. For a simple static set-up of a sphere in $4D$, we have a total set of boundary conditions,
\be\label{eq.expex}
\partial_rg_{tt}\bigg|_{r=r_o}=-\frac{2mG}{r_o^2}\,,\qquad \lim_{r\rightarrow\infty}\partial_r g_{tt}(x^\rho)=0\,,
\ee
in spherical coordinates, in the reference frame of an observer (infinitely) far away from the massive boundary condition at $r=r_o$, with $\lim_{r\rightarrow\infty}g_{tt}=-1$. Here, $m$ is the mass of the object enclosed by the boundary, and $G$ is the gravitational constant. The set-up is of course equivalent to that of a spherical massive object in GR. The relevant point here is what the boundary conditions are characterised by in the present model for spacetime emergence. 

Finally, note that we in this section work under the assumption that an effective spacetime theory has formed. The formulae above are contingent upon that, and do not say anything about where the effective spacetime breaks down, e.g. in the presence of a black hole.

\subsubsection*{Relative movement of boundaries}
As presented in \S\ref{s.qmod}, the effective quantum model includes interactions between quantum fluctuations that in the sense of the emergent spacetime are close to each other. An exchange of momentum is included among the physical processes that can take place at an interaction. Each quantum fluctuation is characterised by a momentum probability distribution around a value $P^\mu_o$ relative to the reference frame, and a small spacetime region is characterised by $\langle P_o^\mu\rangle$, where the average is over the spacetime region (same as for the metric). The propagation of and the exchange of momentum between the quantum fluctuations infer a local equilibration in the momentum. On average, the quantum fluctuations move as a unit, relative to the reference frame.

In the presence of a boundary, the quantum fluctuations adjacent to it will interact with the boundary, and be in equilibrium with it in the same sense they would equilibrate with adjacent quantum fluctuations. This means that the $\langle P_o^\mu\rangle$ of the spacetime region adjacent to a boundary will be in one-to-one correspondence with the velocity of the boundary. Away from the immediate vicinity of the boundary, the standard interactions between the quantum fluctuations give rise to the spacetime configuration, which consequently will be characterised by $R_{\mu\nu}=0$ --- or by a vanishing Riemann tensor, in the case of an absence of massive boundary conditions.

As a consequence of this, if two boundaries move relative to each other, the quantum fluctuations adjacent to each of the boundaries will move with the boundary, and from the perspective of the reference frame, the spacetime will appear to be dragged along by at least one of the boundaries. This is how frame-dragging arises from the effective quantum model in \S\ref{s.qmod}. Frame-dragging is of course well-known from GR, where it e.g. is present in the Kerr geometry and near rotating massive objects like planets; the relevant point here is that (and how) it arises from the effective quantum model in \S\ref{s.qmod}. From the perspective of the reference frame, the frame-dragging arises from that momentum is transferred from the boundary to the quantum fluctuations when the quantum fluctuations interact with the boundary. The momentum is then distributed through the spacetime when the quantum fluctuations interact with each other.

As apparent from the observations above, a second key characteristic of a boundary condition (be it massive or not) is its velocity relative to the reference frame. However, the {\it physical} features arise when multiple boundaries are characterised by non-trivial movement relative to one another. Otherwise, the movement can be rendered trivial by a change of reference frame. In comparison, the Schwarzschild solution is different from the Kerr geometry not only because the Kerr metric is not spherically symmetric; a key feature is that the degree of rotation in the Kerr geometry is set relative to a static, non-rotating, asymptotically flat spacetime (at $r\rightarrow \infty$, in spherical coordinates).

To set up a moving boundary condition in the quantum model at hand is non-trivial, since a non-zero velocity (that is physical in the sense noted above) means that the boundary will be subject to forces, and can deform to a suitable shape. A general, massive object that is spherical when it is at rest cannot be assumed to remain perfectly spherical once a rotation has been added. Disregarding the actual shape of the boundary, the general principles still hold: each boundary has both a massive condition and a velocity condition. If we build on our previous $4D$ example in \eqref{eq.expex}, the boundary condition at $r\rightarrow\infty$ remains the same, but instead of the sphere at the centre of the spacetime, there is a spatial surface $S$ such that
\be\label{eq.mov}
(dx^t)^2 dx^a\partial_a \tilde g_{tt}\bigg|_S=const\leq0\,,\quad \text{in the co-moving reference frame,}
\ee
where $\tilde g_{\mu\nu}$ is the diagonal {\it matrix} that can be obtained from $g_{\mu\nu}$. The co-rotating reference frame encodes the relative velocity of the boundary, which can vary over the surface. Again, this example is just to illustrate what the boundary conditions effectively are characterised by in the effective quantum model at hand. The description is rather general, but captures known properties of GR. The key features of the boundary condition in \eqref{eq.mov} are that it specifies a surface $S$ with a Neumann boundary condition for a presence of mass, and with a velocity relative to the reference frame. This specifies the gradient for the interaction rate at the boundary, and the momentum $\langle P_o^\mu\rangle$ of the adjacent quantum fluctuations. Away from the boundary, the spacetime metric is set by the standard interactions between the quantum fluctuations, and it is Ricci-flat.

\section{Summary and outlook}
We have presented an effective quantum model, built on a conjectured (but possible) effective behaviour of quantum fluctuations in the vacuum, which in the large-scale limit gives GR in vacuum regions as an effective theory. We have illustrated \emph{that} and \emph{how} GR can emerge in a way similar to how hydrodynamics arises from microscopic dynamics; a scenario that should be taken into consideration. The key observations are that: {\it(i)} Gaussian distributions, whose shape in spacetime is set relative to the incoming interactions from the surroundings, can be used to define an internal dynamics which builds up a spacetime, with a metric. {\it (ii)} When the distributions furthermore perform a random walk through the spacetime, the step combination (a combination of a contribution from the particle momentum, and a randomly generated step from the Gaussian distribution) can be used to obtain $R_{\mu\nu}=0$, at the same time as the expected spacetime position of the distribution is forced to obey the geodesic equation.

The advantage of the model is that it constitutes a simple way to get GR as an effective theory at large scales, from effective properties at the quantum scale. The interesting thing will be if the model can be used to connect detailed quantum interaction properties to GR. This would require a more detailed analysis of the interactions at the quantum level, i.e. of what is required of them for the effective model properties to arise. It would also be relevant to compare those detailed interaction properties with the quantum interactions present in gauge theories with gravity duals, such as SYK.

Moreover, we have only discussed the case with $R_{\mu\nu}=0$ in detail. It would be relevant to extend the effective quantum model to include $R_{\mu\nu}\neq0$ and $R\neq0$, the first which is discussed a bit in the appendix. An $R\neq0$ scenario in regions only supported by the vacuum would seem to imply a change in interaction frequency of the quantum fluctuations. A suggestive picture is that a loss of interaction frequency (with the spacetime components effectively gliding apart) would correspond to an effective expansion of the spacetime.

For a full picture of the physics, it would also be relevant to look at the quantum interactions in detail, to determine what happens when the interactions between the quantum fluctuations cease to be stable and frequent enough to give rise to a GR spacetime at large scales.

\appendix

\section{At the level of the quantum interactions}\label{s.lqi}
In addition to the effective quantum model discussed in the main text, it is also relevant to consider the physics at the level of the quantum interactions. This we do in some detail below. Although this discussion is not comprehensive, it concerns some key properties of the quantum interactions, when the interactions are viewed from the perspective of individual interactions.
\\\\
Based on the effective quantum model described in \S\ref{s.eqmod}, it is possible to identify some key features that the quantum interactions need to be characterised by. These requirements, that are necessary for the effective quantum model to manifest itself, fall into the categories of {\it (i)} fundamental requirements on the interactions and the quantum fluctuations, and {\it (ii)} requirements for the effective theory to be a cohesive spacetime. When the requirements in the first category are fulfilled, the second category includes scenarios that are of relevance to understand `spacetime' in regions where the effective theory (GR) breaks down. Here, we will discuss the conditions {\it (i-ii)} a bit qualitatively. They represent topics that need thorough analysis in themselves, and the focus of the main text is to show a scenario for how quantum processes in the spacetime can give rise to GR in vacuum regions (for which the effective quantum model is sufficient), not to provide a full presentation of the small-scale physics.

The quantum fluctuations must be characterised by physical properties sensitive to $S^{d-1}$ and the time directions. With respect to the time directions, this means that the particle must interact with other particles, when those are available, thus creating an `event log'. With respect to the spatial directions, an example was given in \S\ref{s.eqmod}: spin 1/2 particles posses the right characteristics for $D=(1+3d)$ spacetimes, since the spin orientation covers $S^2$. The physical property of the spin, which in spacetime is given relative to an $S^2$, equivalently has an origin in the way the particle interacts. In the current ansatz, that interaction property provides a definition of spatial orientation. In addition, the interactions between the particles must in some sense be delayed, and/or the strength of their impact reduced, in a way compatible with an effective presence of spatial distance.

The presence of a reference frame $\{x^\mu\}$ requires quite a bit more of the quantum fluctuations. For a cohesive spacetime to form, there must be some effective memory and communication of the interaction history of a particle, upon interaction. For example, if particle $A$ interacts with particle $B$, and after that with particle $C$, there must be an increased likelihood for interaction between particles $B$ and $C$. In some sense, their mutual interaction with the same particle must establish them to be `nearby' each other. The shared information established at interaction must also be retained. For example, if an interaction has provided a relative orientation between the particles (relative to the $S^{d-1}$s of the particles), that orientation can only change gradually, as the particles continue with interacting with other particles (moving away in a random walk).\footnote{This process has parallels to particles entangling at interaction, and then slowly disentangling as they interact with other particles. A relative orientation of the $S^{d-1}$ frames could e.g. be modelled through the spin orientation model analysed in \cite{Karlsson:2019vpr,Karlsson:2019vlf}, which basically assigns an individual coordinate system to each particle, albeit without a concept of distance.} Otherwise, no cohesive sense of direction can form. Furthermore, the information on the interaction history, which takes different paths after two particles have interacted, must intersect at a later point in the event log for a relative time to be established between any two particles. This likely also provides a feedback in the system, reinforcing the spacetime structure, and making sure that there is a continuous connection between any two points in spacetime. Overall, it is relevant to note that for a cohesive spacetime to form, the interaction frequency in some sense also must be high enough --- so that clusters of particles can form, where the $S^{d-1}$ frames of the particles remain quite fixed relative to one another.

While a full account of the quantum interactions at the level of individual interactions would require more detail than what is given here, an apparent consequence of the interaction origin ansatz is that when interactions no longer connect two points in spacetime, there is nothing to sustain the formation of an effective spacetime that can be described relative to a reference frame $\{x^\mu\}$. In GR, this would imply that the effective theory would break down in regions that information (or equivalently, light) cannot pass through or away from. This could provide a mechanism for new physics near black hole event horizons, implying a phase transition away from the effective theory (GR, a phase characterised by an ordered spacetime) to a phase dominated by quantum statistical fluctuations. Other scenarios of this kind include the fuzzball proposal \cite{Mathur:2005zp} and the suggestion in \cite{Chapline:2000en} for a quantum phase transition at black hole event horizons.

\subsubsection*{The role of the spacetime dimensions}
In this model, time is a list for the order of events. Each interaction between two particles represents a shared event which takes place in a series of events. This can be depicted in different ways, but one useful way is in terms of time, i.e. to assign an average interaction rate to each particle (a frequency), and to describe the event log in terms of time. This way of assigning time to interactions also has parallels in how time currently is defined using atomic clocks.

The spatial dimensions are defined through the interaction properties of the quantum fluctuations. Orientation is given by the $S^{d-1}$ physical properties, provided that they can be sufficiently correlated between several (preferably very many) quantum fluctuations. Spatial distance and causality are a bit less tangible in the model. The spatial distance is given by $c \delta t$ with $\delta t$ denoting a time delay when information is communicated between or sent from one quantum particle to another, at the speed of light and at a corresponding angle in the $S^{D-1}$ frame. Causality in the large-scale limit would impose certain restrictions on the interaction properties; we briefly mention one such property in \S\ref{s.Gprob}: the probability for a particle to take a step back in time. For more details on how the dimensions can emerge, we refer to \cite{Karlsson:2019avt}.

\section{Sketched example: electric field}\label{s.efield}
Regions of spacetime that only contain quantum fluctuations can support more configurations than those typically called vacuum regions of spacetime, i.e. regions with $R_{\mu\nu}=0$. The most straightforward example is when a boundary condition includes electric change, and an electric field extends through the spacetime. The presence of the field makes the configuration deviate from the vacuum configuration, through $R_{\mu\nu}\neq0$, but the only particles present remain those present in a typical vacuum region. Since our ansatz for spacetime emergence is built on interactions between quantum fluctuations, it should be able to encompass all spacetime configurations that are supported solely by quantum fluctuations, including the presence of electromagnetic fields and a non-zero cosmological constant. Moreover, it is quite straightforward to identify the mechanisms behind how $R_{\mu\nu}\neq0$ works out in our model, in the presence of a boundary condition with electric charge.

In the effective quantum model in \S\ref{s.qmod}, the spacetime configuration is determined by the average behaviour of the quantum fluctuations. Provided that the quantum fluctuations interact frequently enough, an effective spacetime emerges at large scales, and the behaviour of the quantum fluctuations can be interpreted relative to a reference frame. From that perspective, each of the quantum fluctuations interact quite locally. A reference frame can always be chosen so that the interaction rate with the surrounding quantum fluctuations has a Gaussian fall-off. Consequently, for an electric field to {\it effectively} be present, it must exist through a perturbed pattern of interaction between the quantum fluctuations, so that the effect of the field can be carried away, far from the boundary that has the electric charge, in a chain reaction.

The simple solution to what that chain reaction is, is the following. The quantum fluctuations include both charged and neutral particles. The charged particles close to the boundary with the electric charge will be affected by the charged boundary condition. On average, half of the charged particles will be repelled by the boundary, and the others will be attracted by it. Effectively, two opposite flows will be instated, locally and perpendicular to the boundary. The spacetime region will still be neutral, but two flows will be present. Further away from the boundary, the quantum fluctuations will not directly interact with the boundary condition. Instead, a quantum fluctuation will interact with the surrounding quantum fluctuations, and the two flows of charged particles will mean that the interaction frequencies of the moving particles will be subject to a Doppler shift relative to the quantum fluctuation. Effectively, the average frequency at which those particles interact with the quantum fluctuation will deviate from their real interaction frequency. The shift in effective interaction rate will mean that particles of one type of charge (positive/negative) moving towards the quantum fluctuation will interact more frequently with the particle, while particles of the other type (moving away) will interact less frequently with it, and the surroundings will appear charged to the quantum fluctuation, despite that it is not. Consequently, a charged quantum fluctuation will be induced to move accordingly, i.e. to take part in the flow, either towards or away from the boundary condition (depending on its type of charge). Effectively, this will create a chain reaction away from the boundary that propagates through the spacetime.

Regarding $R_{\mu\nu}\neq0$, the two parallel flows of charged quantum fluctuations mean that charged quantum fluctuations are subject to a force, and a separate process for particle propagation is introduced. For that reason, \eqref{eq.pProp} is no longer is enforced, and the spacetime can have $R_{\mu\nu}\neq0$. We will not treat the form of the resultant $R_{\mu\nu}$ in detail here. $R_{\mu\nu}\propto T_{\mu\nu}$, where $T_{\mu\nu}$ is the stress-energy tensor of the electric field, is however very plausible since the deviation from the condition in \eqref{eq.pProp} is explicitly caused by the presence of an effective electric field, carried by the quantum fluctuations. In a bit more detail, it is however possible to observe that the Ricci tensor must depend on the electric field effectively present at each point in spacetime $x_o^\rho$, and that $R_{\mu\nu}$ must be independent of the type of charge (positive/negative) present at the boundary. This follows from that the effect on the spacetime metric is caused by the same process (two opposite flows of charged particles) regardless of the type of charge specified by the boundary. Consequently, the $R_{\mu\nu}$ must depend on some function of $E^2$, as is the case for the stress-energy tensor of an electric field. A property of $R\neq0$ can also be regarded as motivated from that the electric field merely introduces a preferred direction of propagation in the spacetime, to the quantum fluctuations. No energy is introduced/removed at points in the spacetime (away from the boundary), which should correspond to a special case of spacetime, i.e. configurations with $R\neq0$. Moreover, the electric field present in the spacetime, $E^i(x^\rho)$, will depend on the configuration of the boundary, and be set by the requirement that the field is divergence free. From the perspective of the spacetime components, that property arises from that the quantum fluctuations only carry an image of the charge present at the boundary. There is no net flow across the surface of any spacetime region containing only quantum fluctuations. Equivalently, the only charge present is at the boundary, and the electric field outside the boundary must be divergence free.

\providecommand{\href}[2]{#2}\begingroup\raggedright\endgroup

\end{document}